\documentclass[prl,twocolumn,superscriptaddress,showpacs]{revtex4}
\usepackage{graphicx}

\begin{document}

\title{Superlight small bipolarons in the presence of strong Coulomb repulsion}

\author{J.P. Hague}
\affiliation{Department
 of Physics, Loughborough University, Loughborough, LE11 3TU, United Kingdom}

\author{P.E. Kornilovitch}
\affiliation{Hewlett-Packard Company, 1000 NE Circle Blvd,
Corvallis, Oregon 97330, USA}

\author{J.H. Samson}
\affiliation{Department
 of Physics, Loughborough University, Loughborough, LE11 3TU, United Kingdom}

\author{A.S. Alexandrov}
\affiliation{Department
 of Physics, Loughborough University, Loughborough, LE11 3TU, United Kingdom}

\date{June 1, 2006}

\begin{abstract}

We study a lattice bipolaron on a staggered triangular ladder and
triangular and  hexagonal lattices   with both long-range
electron-phonon interaction and strong Coulomb repulsion using a novel
continuous-time quantum Monte-Carlo (CTQMC) algorithm extended  to the
Coulomb-Fr\"ohlich model with two particles. The algorithm is
preceded by an exact integration over phonon degrees of freedom, and
as such is extremely efficient. The bipolaron effective mass and
bipolaron radius are computed. Lattice  bipolarons on such lattices have a novel crablike motion, and
are small but very light in a wide range of parameters, which leads
to a high Bose-Einstein condensation temperature. We discuss the relevance of our
results with current experiments on cuprate high-temperature
superconductors and propose a route to room
 temperature superconductivity.


\end{abstract}

\maketitle

As recognized by Landau, Pekar and Fr\"ohlich, an electron may drag
a lattice distortion as it moves through an ionic material, leading
to a new particle - the polaron, which has quite different
properties to the original electron (for reviews see, for example,
Refs. \cite{ref1,ref2}). At weak coupling, two polarons can be bound
into a large bipolaron via exchange forces, without assuming
anything more complicated than the Fr\"ohlich electron-phonon
interaction \cite{ref9}. On increasing density large bipolarons
overlap, giving rise to either a conventional (BCS) superconductor
or a normal metal. Electron-phonon interactions may 
overcome the Coulomb repulsion between
electrons, so the resulting interaction becomes attractive at a
distance of the order of the lattice constant \cite{ref10}. Then two
small polarons form tightly bound pairs, i.e. small bipolarons, in
the strong electron-phonon coupling limit. Earlier studies
\cite{ref11} considered small bipolarons as entirely localized
objects. However, a perturbation expansion with respect to the
hopping integral has proved they are itinerant quasiparticles
existing in Bloch states for any finite coupling with phonons and
forming a Bose-Einstein condensate (BEC) of charge $2e$  bosons at
sufficiently low temperatures \cite{ref12}.

For very strong electron-phonon coupling, polarons become
self-trapped on a single lattice site. The energy of the resulting
small polaron is given as $E_p=-\lambda zt$, where $\lambda$ is the
electron-phonon coupling constant, $t$ is the hopping parameter and
$z$ is the coordination number. Expanding about the atomic limit in
small $t$ (which is small compared to $E_p$ in the small polaron
regime, $\lambda>1$)
 the polaron mass is
computed as $m^{*}=m_0\exp(\gamma z\lambda/\hbar\omega)$ , where
$\omega$ is the frequency of Einstein phonons, $m_0$ is the rigid lattice
band mass, and $\gamma$ is a numerical constant.
For the Holstein model \cite{ref14}, which is purely site local,
$\gamma=1$. Bipolarons are on-site singlets in the Holstein model
and their mass $m_{H}^{**}$ appears only in the second order of $t$
\cite{ref12} scaling as $m_{H}^{**}\propto (m^{*})^2$ in the limit
$\hbar\omega\gg\Delta$ , and as $m_{H}^{**}\propto(m^*)^{4}$ in a
more realistic regime $\hbar\omega\ll\Delta$ \cite{ref10}. Here
$\Delta=2E_p-U$ is the bipolaron binding energy, and $U$ is the
on-site (Hubbard) repulsion. Since the Hubbard $U$ is about 1 eV or
larger in strongly correlated materials, the electron-phonon
coupling must be large to stabilize on-site bipolarons and the
Holstein bipolaron mass appears very large, $m_{H}^{**}/m_0>1000$,
for realistic values of phonon frequency.

 This estimate led some authors to the conclusion that the formation of
itinerant small polarons and bipolarons in real materials is
unlikely \cite{ref15}, and high-temperature bipolaronic
superconductivity is impossible \cite{ref16}. However, one should
note that the Holstein model is an extreme polaron model, and
typically yields the highest possible value of the (bi)polaron mass
in the strong coupling limit.  Many advanced materials with low
density of free carriers and poor mobility (at least in one
direction) are characterized by poor screening of high-frequency
optical phonons and are more appropriately described by a long-range
Fr\"ohlich electron-phonon interaction \cite{ref10}. For a
long-range Fr\"ohlich interaction the parameter $\gamma$ is less
than 1 ($\gamma\approx 0.3$ on the square lattice and $\gamma\approx
0.2$ on the triangular lattice \cite{ref17,ref19}), reflecting the
fact that in a hopping event the lattice deformation is partially
pre-existent. Hence the unscreened Fr\"ohlich electron-phonon
interaction provides relatively light small polarons, which are
several orders of magnitude lighter than small Holstein polarons.
This has been confirmed by numerical Monte-Carlo simulations
\cite{ref20,ref19}, Lanczos diagonalization \cite{ref21} and
variational calculations \cite{ref22}.

This unscreened Fr\"ohlich  interaction combined with on-site
repulsive correlations can also bind holes into intersite mobile
bipolarons \cite{ref17,ref23}. Using an advanced variational method
Bon\v{c}a and Trugman \cite{ref22} studied the chain model of Ref.
\cite{ref20} with two electrons for nearest-neighbor e-ph
interaction and a Hubbard $U$. Intersite bipolarons of Ref.
\cite{ref22} propagate along the chain with a mass which is still of
the second order in the polaron mass as in the Holstein model.

Here we study a bipolaron on a staggered triangular ladder (1D),
triangular (2D) and strongly anisotropic hexagonal (3D) lattices using a
continuous-time quantum Monte-Carlo technique.  On such lattices,
bipolarons are found to move with a crab like motion (Fig. 1(b)), which is
distinct from the crawler motion (Fig. 1(c)) found on cubic lattices
\cite{ref12,ref22}. Such bipolarons are small but very light for a
wide range of electron-phonon couplings and phonon frequencies.

We use a generic Coulomb-Fr\"{o}hlich model of electron-phonon
interactions which has the following Hamiltonian,
\begin{eqnarray}
H & = & - t \sum_{\langle \mathbf{nn'} \rangle\sigma}
c^{\dagger}_{\mathbf{n'}\sigma} c_{\mathbf{n}\sigma} + \sum_{
\mathbf{nn'}\sigma}V(\mathbf{n},\mathbf{n}') c^{\dagger}_{\mathbf{n}\sigma}
c_{\mathbf{n}\sigma}c^{\dagger}_{\mathbf{n'}\bar{\sigma}} c_{\mathbf{n'}\bar{\sigma}} \nonumber\\ &+& 
\sum_{\mathbf{m}} \frac{\hat{P}^{2}_\mathbf{m}}{2M} + 
\sum_{\mathbf{m}} \frac{\xi^{2}_{\mathbf{m}} M\omega^2}{2} -
\sum_{\mathbf{n}\mathbf{m}\sigma} f_{\mathbf{m}}(\mathbf{n})
c^{\dagger}_{\mathbf{n}\sigma} c_{\mathbf{n}\sigma} \xi_{\mathbf{m}}\nonumber
\: . \label{eq:four}
\end{eqnarray}
%
%
 Each vibrating ion has one phonon degree of freedom $\xi_\mathbf{m}$  associated with a single atom.
The sites are numbered by the indices $\mathbf{n}$ or $\mathbf{m}$
for electrons and ions respectively. Operators $c$ annihilate
electrons. The phonon subsystem is a set of independent oscillators with
frequency $\omega$ and mass $M$. Here
$\langle\mathbf{n}\mathbf{n}'\rangle$ denote pairs of nearest
neighbors, and $\hat{P}_{\mathbf{m}}=-i\hbar\partial/\partial\xi_{\mathbf{m}}$
is the ion momentum operator. Coulomb repulsion
$V(\mathbf{n}-\mathbf{n}')$ is screened up to the first nearest
neighbors, with on site repulsion $U$ and nearest-neighbor
repulsion $V_C$. In contrast, the Fr\"ohlich interaction is assumed
to be long-range, due to unscreened interaction with c-axis
high-frequency phonons \cite{ref10}. The form of the interaction
with c-axis polarized phonons is specified via the force
function\cite{ref20},
$f_{\mathbf{m}}(\mathbf{n})=\kappa\left[(\mathbf{m}-\mathbf{n})^2+1\right]^{-3/2}$,
where $\kappa$ is a constant. The dimensionless electron-phonon
coupling constant $\lambda$ is defined as
$\lambda=\sum_{\mathbf{m}}f^{2}_{\mathbf{m}}(0)/2M\omega^2 zt$ which
is the ratio of the polaron energy at $t=0$  to the kinetic energy
of the free electron $zt$.

In the limit of high phonon frequency $\hbar\omega\gg t$ and large
on-site Coulomb repulsion, the model is reduced to an extended
Hubbard model with intersite attraction and suppressed
double-occupancy \cite{ref23}. Then
the Hamiltonian can be projected onto the subspace of nearest
neighbor intersite crab bipolarons. In contrast with the crawler bipolaron, the crab
bipolaron's mass scales linearly with the polaron mass
($m^{**}=4m^*$ on the staggered chain and $m^{**}=6m^{*}$ on the triangular lattice).  Here we
formulate the following question: Can such a bipolaron exist for
more realistic intermediate values of the electron-phonon coupling
and phonon frequency?

To answer this question, we have extended the CTQMC algorithm
\cite{ref24,ref20,ref25,ref19} to systems of two particles with
strong electron-phonon interactions.  We have solved the bipolaron
problem on a staggered ladder (Fig.1), triangular and anisotropic hexagonal
lattices from weak to strong coupling in a realistic parameter range
where usual limiting approximations fail.

The CTQMC method employed here has been described in detail with
regard to the single polaron problem in Refs.
\cite{ref24,ref25,ref19}. Here we give a quick overview of the
extended algorithm.  The initial step is to determine the effective
bipolaron action that results when the phonon degrees of freedom
have been integrated out analytically.  The action is a functional
of two polaron paths in imaginary time  which form the bipolaron and
is given by the following double integral,
 \begin{widetext}
\begin{eqnarray}
A[{\bf r}(\tau)] & = & \frac{z\lambda\bar{\omega}}{2\Phi_0(0,0)}
\int_0^{\bar\beta} \int_0^{\bar\beta} d \tau d \tau'
e^{-\bar{\omega} \bar\beta/2}\left( e^{\bar{\omega}(\bar\beta/2-|\tau-\tau'|)} +
                               e^{-\bar{\omega}(\bar\beta/2-|\tau-\tau'|)} \right)
\sum_{ij}\Phi_0[\mathbf{r}_i(\tau),\mathbf{r}_j(\tau')]  \\
 & + & \frac{z\lambda\bar{\omega}}{\Phi_0(0,0)}
 \int_0^{\bar\beta} \int_0^{\bar\beta} d \tau d \tau' e^{- \bar{\omega} \tau}
 e^{-\bar{\omega}( \bar\beta - \tau')}
 \sum_{ij}\left( \Phi_{\Delta\mathbf{r}}[\mathbf{r}_i(\tau),\mathbf{r}_j(\tau')] -
 \Phi_0[\mathbf{r}_i(\tau),\mathbf{r}_j(\tau')]\right) -\int_0^{\beta}V(\mathbf{r}_1(\tau),\mathbf{r}_2(\tau))\,d\tau \: .\nonumber
\label{eq:seven}
\end{eqnarray}
\end{widetext}
The full interaction between the particles is
$\Phi_{\Delta\mathbf{r}}[\mathbf{r}(\tau),
\mathbf{r}(\tau')]=\sum_{\mathbf{m}}f_{\mathbf{m}}[\mathbf{r}(\tau)]f_{\mathbf{m}+\Delta\mathbf{r}}[\mathbf{r}(\tau')]$
where the vector $\Delta\mathbf{r}=\mathbf{r}(\beta)-\mathbf{r}(0)$
is the difference between the end points of one of the paths in the
non-exchanged configuration (here $\bar{\omega}=\hbar \omega/t$  and
$\bar{\beta}=t/k_BT$).  The indices $i=1,2$ and $j=1,2$ represent
the fermion paths. $V(\mathbf{r}_1,\mathbf{r}_2)$ is an
instantaneous Coulomb repulsion. From this starting point, the
bipolaron is simulated using the Metropolis Monte-Carlo (MC) method. The
electron paths are continuous in time with hopping events (or kinks)
introduced or removed from the path with each MC step.
Analytic integration is performed over sections of parallel paths.
The ends of the two paths at $\tau=0$ and
$\tau=\beta$ are related by an arbitrary translation, $\Delta\mathbf{r}$. In contrast to the one-particle case, the fixing
of the end configurations limits the update procedure to inserting
and removing pairs of kinks and antikinks in the following ways: (a) addition/removal of two kinks to/from  different paths, (b) addition/removal of a kink-antikink pair to/from one path, (c) addition and removal of a kink to/from a single path (kink shift), (d) kink addition to one path and antikink removal from the other path. On kink insertion/removal, either the top or bottom of the path is shifted, which allows the interparticle distance to change. Another significant difference to the one-particle problem is that the paths can now be exchanged. There are two ways to carry out the exchanges: (1) Inserting/removing multiple kinks/antikinks, or (2) If there is a common segment, one may break the paths at that segment, and splice the bottom half of path 1 to the top half of path 2 and vice-versa. In the exchanged state, updates (a) to (d) with idential shifts for both single updates are combined with (e) Addition of kink and antikink on different paths
(f) Kink addition to 1 path and kink removal from the other
(g) Addition of kink and removal of antikink on same path
(h) Addition/removal of kink pair to one path. The shift types are opposed for these binary update parts, and allow for a change in the interparticle distance in the exchanged configuration, and as such are essential to sample the full configuration space. 
From the ensemble the
ground state bipolaron energy and effective mass are computed as in
Ref.  \cite{ref24}. Also, the bipolaron radius is computed as
$R_{bp}=\left\langle\sqrt{\frac{1}{\beta}\int_{0}^{\beta}(\mathbf{r}_1(\tau)-\mathbf{r}_2(\tau))^2
d\tau}\right\rangle$ , where $\beta\gg(\hbar\omega)^{-1}$.

\begin{figure}
\includegraphics[width=70mm]{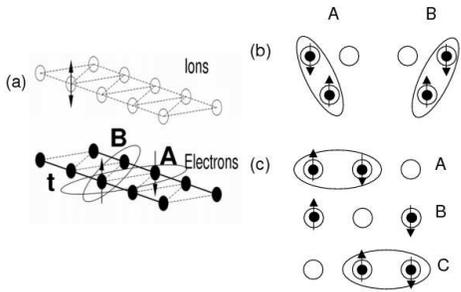}
\caption{(a) Schematic of the ladder model. Electrons sit on
opposite sides (legs) of a staggered ladder with intersite distance
$a$, with ions vibrating across the ladder on an identical system
sitting a height $a$ above the electron legs.  (b) Schematic motion
of the crab bipolaron - the two states are degenerate. (c) Schematic
of the crawler motion. }
\end{figure}

\begin{figure}
\includegraphics[height=75mm,angle=270]{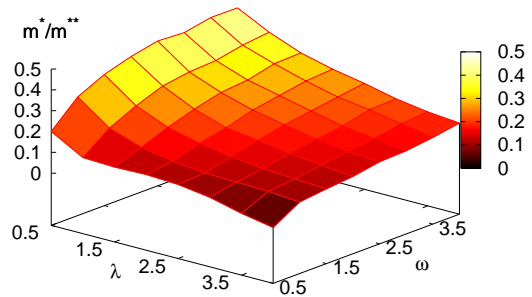}
\caption{(color online) Polaron to bipolaron mass ratio for a range of
$\bar{\omega}$ and $\lambda$ on the staggered ladder.  Mobile small
bipolarons are seen even in the adiabatic regime $\bar{\omega}=0.5$
for couplings $\lambda$ up to 2.5.}
\end{figure}

\begin{figure}
\includegraphics[height=75mm,angle=270]{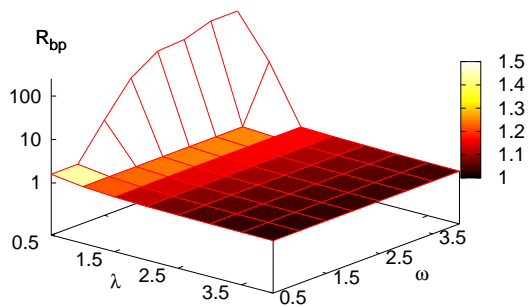}
\caption{(color online) Bipolaron radius (in units of $a$) for a range of
$\bar{\omega}$ and $\lambda$ on the staggered ladder.}
\end{figure}

\begin{figure}
\includegraphics[height=75mm,angle=270]{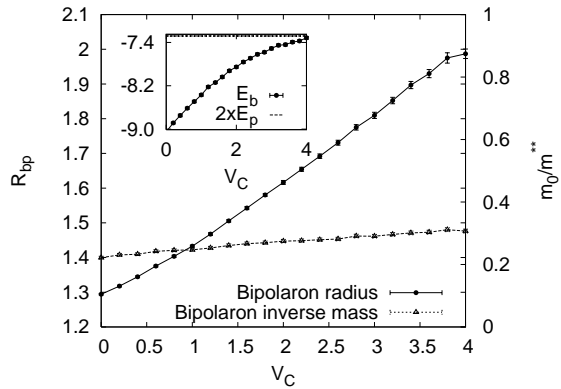}
\caption{Variation of bipolaron energy, mass and radius (in units of
$a$) as intersite Coulomb repulsion $V_C$ is increased for $\lambda$
and $\bar{\omega}=1$ on the staggered ladder.}
\end{figure}

Figure 2 shows the ratio of the polaron to bipolaron masses on the
staggered ladder as a function of effective coupling and phonon
frequency for $V_C=0$. The bipolaron to polaron mass ratio is about
2 in the weak coupling regime ($\lambda\ll1$) as it should be for a
large bipolaron \cite{ref9}. In the strong-coupling, large phonon
frequency limit the mass ratio approaches 4, in agreement with
strong-coupling arguments given above.  In a wide region of
parameter space, we find a bipolaron/polaron mass ratio of between 2
and 4 and a bipolaron radius similar to the lattice spacing, see
Figs. 3 and 4. Thus the bipolaron is small and light at the same
time. Taking into account additional intersite Coulomb repulsion
$V_C$ does not change this conclusion. The bipolaron is stable for
$V_C<4t$ , see Fig. 4 (inset). As $V_C$ increases the bipolaron mass
decreases but the radius remains small, at about 2 lattice spacings.
Importantly, the absolute value of the small bipolaron mass is only
about 4 times of the bare electron mass $m_0$, for
$\lambda=\hbar\omega/t=1$ (see Fig. 4).

The toy problem on the triangular ladder  contains the essential physics of the crab bipolaron. We demonstrate this by simulating the
bipolaron on an infinite triangular lattice including exchanges and
large on-site Hubbard repulsion $U = 20t$. A moderate coupling
$\lambda= 0.5$ and a large phonon frequency $\omega= 2t$ lead to
$m^{**}_{xy}=(3.77\pm0.04)m_{0xy}$ and a small bipolaron radius of
$(2.056\pm0.004)a$. For the triangular lattice,
$m_{0xy}=\hbar^{2}/3a^{2}t$. Finally, we have simulated the
bipolaron in a hexagonal lattice, with out-of-plane hopping
$t'=0.3t$. We have calculated values of the bipolaron mass and radius
for experimentally achievable values of the phonon frequency
$\omega=t=200$meV and electron-phonon coupling $\lambda=0.3$.  We
have found a light in-plane mass,
$m_{xy}^{**}=(4.49\pm0.04)m_{0xy}$. Out-of-plane
$m_{z}^{**}=(68.4\pm1)m_{0z}$ is Holstein like, where
$m_{0z}=\hbar^2/2d^2t'$.  The bipolaron radius is
$R_{bp}=(2.60\pm0.03)a$, sitting mainly in the $xy$ plane.

When bipolarons are small and pairs do not overlap, the pairs can
form  a BEC at
$k_BT_{\mathrm{BEC}}=3.31\hbar^2(2n_B/a^2\sqrt{3}d)^{2/3}/(m_{xy}^{**2/3}m_z^{**1/3})$.
If we choose realistic values for the lattice constants of 0.4 nm in
the plane and 0.8 nm out of the plane, and allow the density of
bosons to be $n_B$=0.12 per lattice site, which easily avoids
overlap of pairs, then
 $T_{\mathrm{BEC}}=323$K.
The long-range Fr\"ohlich interaction combined with Coulomb
repulsion might cause clustering of polarons into finite-size
quasi-metallic mesoscopic textures. However analytical \cite{ref23}
and QMC \cite{ref26} studies of  mesoscopic textures with lattice
deformations and Coulomb repulsion show that pairs (i.e. bipolarons)
dominate over phase separation since they effectively repel each
other \cite{ref1}.

Recently, there has been a large revival in quantitative studies of
polarons  owing to  evidence for polaronic effects in
high-temperature superconductors \cite{ref4,ref5}. There are strong
arguments  in favor of 3D bipolaronic BEC in cuprates \cite{ref10}
drawn using parameter-free fitting of experimental $T_c$ with BEC
$T_{c}$ \cite{alekab}, unusual upper critical fields and the
specific heat \cite{ZAV}, and more recently the normal state
diamagnetism \cite{aledia}. Here we have presented the numerically
exact bipolaron mass and size, which put these arguments (which require very light bipolarons in the intermediate coupling, moderate phonon-frequency regime) on a solid
microscopic ground.

 In summary,  the CTQMC algorithm to simulate bipolarons
in the Coulomb-Fr\"ohlich model has been extended, leading to an
unusual bipolaron configuration that is small and superlight. Such a
particle has been found in a wide parameter range using CTQMC in
triangular lattices with achievable phonon frequencies and couplings
in the presence of strong Coulomb repulsion. Such bipolarons  could
easily have a superconducting transition in excess of room
temperature.  We believe that the following recipe is worth
investigating to look for room-temperature superconductivity: (a)
The parent compound should be an ionic insulator with light ions to
form high-frequency optical phonons, (b) The structure should be
quasi two-dimensional to ensure poor screening of high-frequency
c-axis polarized phonons,  (c) A triangular lattice is required in
combination with  strong, on-site
 Coulomb repulsion to form the small superlight Crab
bipolaron  (d) Moderate carrier densities are required to keep the
system of small bipolarons close to the dilute regime.

The authors acknowledge support from EPSRC (UK)
grants no EP/C518365/1 and no EP/D07777X/1, and useful discussions
with J. Devreese, P. Edwards, V. Kabanov, Y. Liang,
M. Stoneham, and P. Zhao.

\end{document}